\documentclass{edp-conf}
\usepackage{graphicx}

\begin{document}

\TitreGlobal{SF2A 2004}

\title{Mrk~421 and 501 above 60~GeV: the Influence of CELESTE's Energy Scale on the Study of Flares and Spectra}
\author{Brion, E.}\address{CENBG, Domaine du Haut-Vigneau, BP 120, 33175 Gradignan Cedex, France\\
E-mail: \texttt{brion@cenbg.in2p3.fr}}
\author{the CELESTE collaboration}
\runningtitle{The Influence of CELESTE's Energy Scale}
\setcounter{page}{237}
\index{Brion, E.}

\maketitle

\begin{abstract}
The CELESTE atmospheric Cherenkov detector ran until June 2004. It has detected several $\gamma$-ray flares of Mrk~421 since 1999. The new data analysis provides better background rejection. We compare our light curves with X-ray data. We significantly improved our understanding of the atmosphere, and of the optical throughput of the detector. This will allow a spectral measurement for Mrk~421 with smaller uncertainties and a more sensitive search for a signal from Mrk~501.
\end{abstract}

%

\section{Introduction}

CELESTE was a Cherenkov experiment using 53 heliostats of the former \'Electricit\'e de France solar plant in the French Pyrenees at the Th\'emis site. Its principle is to detect the Cherenkov light of the electromagnetic shower produced in the atmosphere by the $\gamma$-rays coming from high energy astrophysical sources. The light is reflected to secondary optics and photomultipliers installed at the top of the tower. Finally, it is sampled to be analysed (Par\'e 2002).

We significantly improved our data analysis and have better background rejection since the 2000 status (de Naurois 2002). Changes, described elsewhere in this volume (Manseri 2004), were made in hardware (new heliostats and pointing strategy) and in software (new data analysis). The Crab sensitivity was improved from 2.2~$\sigma/\sqrt{\mathrm{h}}$ to 5.8~$\sigma/\sqrt{\mathrm{h}}$. We present the improved light curves for Mrk~421. We also now better understand the optical throughput of the detector with our new simulation.

\section{Observations of Mrk~421 since 2002} \label{sec:Mrk421}

This stable analysis and good sensitivity provide a good detection of the Mrk~421 flares. A 19~$\sigma$ detection during 10~h since 2002 gives a mean of 5.6~$\gamma/$min. The light curve is presented in figure~\ref{fig:all}~(a).

\section{Simulation improvement} \label{sec:simulation}

\begin{figure}[htbp]
\begin{center}
   \includegraphics[width=6cm]{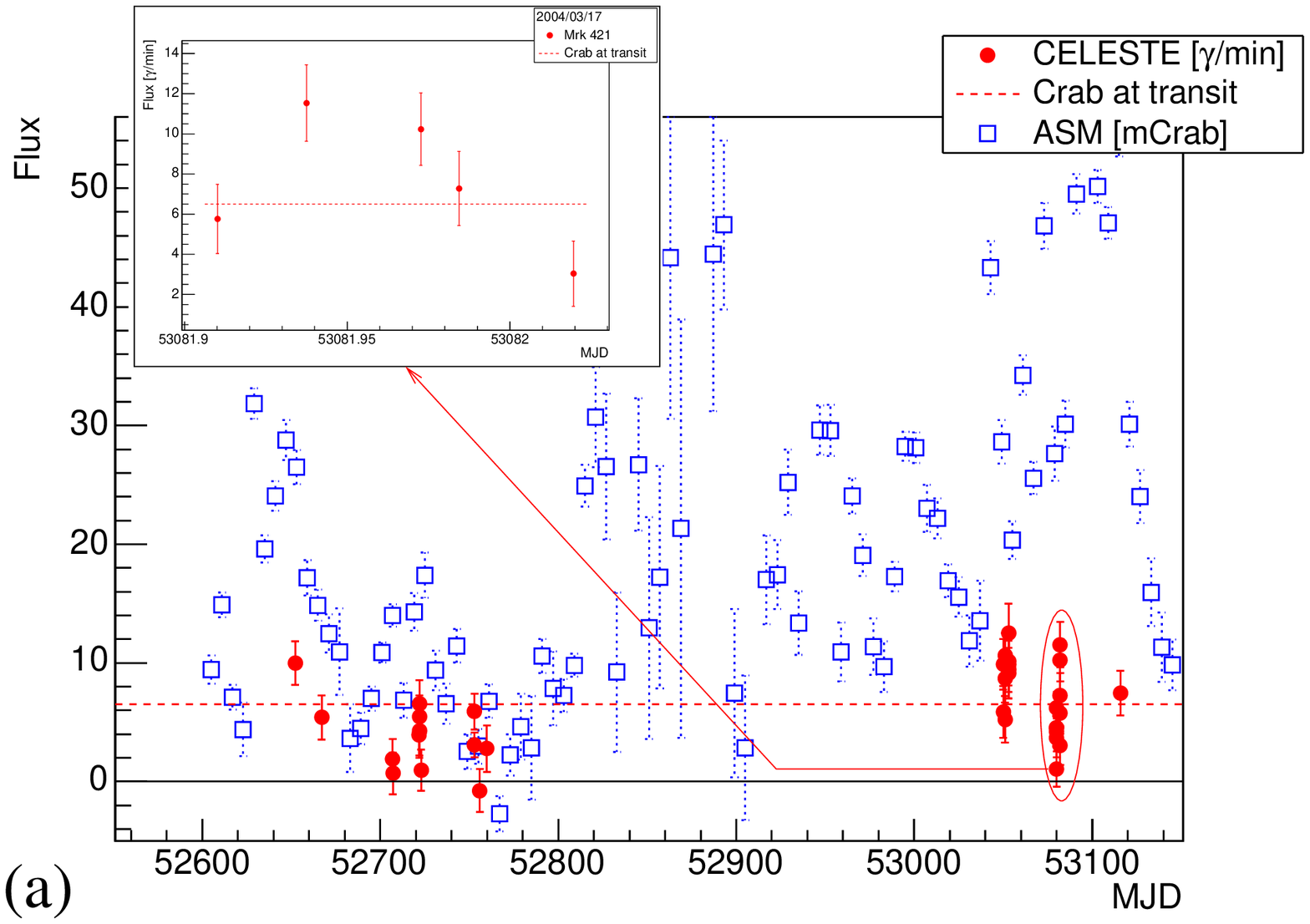}
   \includegraphics[width=6cm]{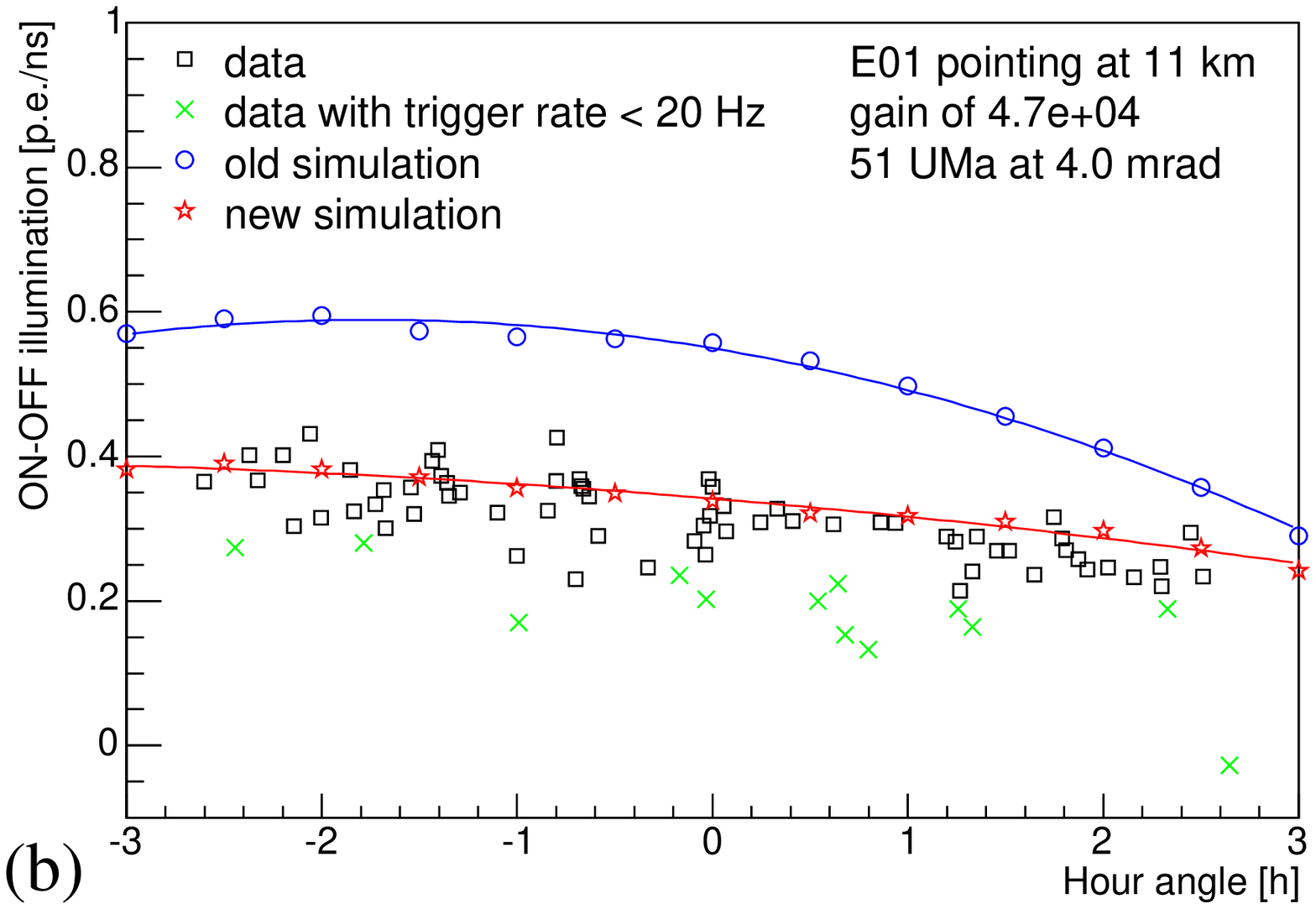}
\end{center}
   \caption{(a) Light curve of Mrk~421 since 2002 seen by CELESTE (red points in $\gamma$/min corrected for hour angle efficiency, 1~point is 20~min data, Crab transit rate shown for reference) and RXTE/ASM (blue squares in mCrab, 6~day bins). The zoom shows the $17^{\mathrm{th}}$ of March 2004 flare. (b) ON$-$OFF illumination from the star 51~UMa (M$_\mathrm{B}=6.16$) in the field of view of Mrk~421 as a function of hour angle: the new simulation with our corrections (red stars) fits the data well (black squares) whereas the old simulation (blue circles) was 50~\% too high. Note that the difference is smaller for $\gamma$-ray showers (an extended light source) than for star (point sources).}
   \label{fig:all}
\end{figure}

To constrain the energy scale of the experiment and have reliable acceptances to deduce a spectral measurement for Mrk~421, the simulation must reproduce the detector response faithfully (Brion 2004). The LIDAR operating on the site provided a better determination of the atmospheric extinction (Buss\'ons Gordo 2004). A stellar photometry study showed that the old simulation was too optimistic and now gives good agreement with the data (mirror reflectivities decreased, nominal focussing degradation, and verification of the photomultiplier gains, see figure~\ref{fig:all}~(b)).\\

Finally, this new analysis will allow an improved investigation for Mrk~501's behaviour since the old analysis gave 2.5~$\sigma$ significance during 14.5~h data.

\vspace{2\baselineskip}


\end{document}